\documentclass[preprint, 5p, times,twocolumn]{elsarticle}

\usepackage{amssymb}

\usepackage{subfig}
\usepackage{tabularx}
\usepackage{booktabs}
\newcommand{\tabitem}{~~\llap{\textbullet}~~}
\usepackage{multirow}
\usepackage[hyphens]{url}

\journal{Microelectronics Journal}

\begin{document}

\begin{frontmatter}


\title{Impact of gate-level clustering on automated system partitioning of 3D-ICs}
\tnotetext[]{Accepted for publication on July 13th 2023: \url{https://doi.org/10.1016/j.mejo.2023.105896}. \textcopyright 2023. This manuscript version is made available under the CC-BY-NC-ND 4.0 license \url{https://creativecommons.org/licenses/by-nc-nd/4.0/}}

\author[beams]{Quentin Delhaye\corref{cor1}}
\ead{quentin.gc.delhaye@ulb.be}
\author[imec]{Eric Beyne}
\author[info]{Joël Goossens}
\author[imec]{Geert Van der Plas}
\author[beams]{Dragomir Milojevic}

\affiliation[beams]{organization={Bio- Electro- and Mechanical Systems, Université libre de Bruxelles},
            addressline={Av. Franklin Rooselvelt 50, CP 165/56}, 
            city={Brussels},
            postcode={1050},
            country={Belgium}}

\affiliation[imec]{organization={imec},
            addressline={Kapeldreef 75}, 
            city={Leuven},
            postcode={3001},
            country={Belgium}}

\affiliation[info]{organization={Computer Science Departement, Université libre de Bruxelles},
            addressline={Boulevard du Triomphe}, 
            city={Brussels},
            postcode={1050},
            country={Belgium}}

\cortext[cor1]{Corresponding author}

\begin{abstract}
When partitioning gate-level netlists using graphs, it is beneficial to cluster gates to reduce the order of the graph and preserve some characteristics of the circuit that the partitioning might degrade.
Gate clustering is even more important for netlist partitioning targeting 3D system integration.  
In this paper, we make the argument that the choice of clustering method for 3D-ICs partitioning is not trivial and deserves careful consideration.
To support our claim, we implemented three clustering methods that were used prior to partitioning two synthetic designs representing two extremes of the circuits medium/long interconnect diversity spectrum.
Automatically partitioned netlists are then placed and routed in 3D to compare the impact of clustering methods on several metrics.
From our experiments, we see that the clustering method indeed has a different impact depending on the design considered and that a circuit-blind, universal partitioning method is not the way to go, with wire-length savings of up to 31\%, total power of up to 22\%, and effective frequency of up to 15\% compared to other methods.
Furthermore, we highlight that 3D-ICs open new opportunities to design systems with a denser interconnect, drastically reducing the design utilization of circuits that would not be considered viable in 2D.
\end{abstract}

\begin{keyword}
System partitionability \sep 3D-ICs \sep clustering \sep mincut partitioning
\end{keyword}

\end{frontmatter}


\section{Introduction \& Overview}\label{sec:intro}

Three-dimensional integrated circuits (3D-ICs) are no longer just a perspective in a More-than-Moore era~\cite{nirmalya_maity_as_2022}.
Alongside stretching current technology nodes, they are readily made available to the general public~\cite{advanced_micro_devices_inc_amd_2022}.
There are still many challenges to overcome~\cite{eric_beyne_3d_2021}, such as tools used for automated 3D design place and route (P\&R) and optimizations, and 3D-IC characterization: power, thermal, IR-drop, electromigration, mechanical analysis, etc.~\cite{subhajit_chatterjee_frequency-scaled_2022, xiaonan_guan_distribution_2023}
To overcome the absence of true 3D physical implementation tools, academia and industry typically trick the existing 2D tools. An example of such 3D flow, based on commercial quality tools, is the ``Die-by-Die'' flow~\cite{sisto_design_2019}, where each die is processed individually, in a sequence.
Such flows have been used to produce encouraging Power, Performance and Area (PPA) metrics of 3D-ICs compared to the original 2D circuits.

A preliminary requirement to enable 3D system design is to partition the circuit, splitting the original netlist in two (or more) separate dies. The partitioning decision, that is which standard cells or memory macro is to be placed on which tier, can be manual or automated. The former approach can be successfully used for functional, coarse grain system partitioning such as Memory-on-Logic. However, finer grain system partitioning enabled with fine 3D interconnect pitch requires automation, e.g. using graphs that represent the interconnect of a circuit. In this case, the following question arises: What can we do to guide the partitioning decision in order to optimize the 3D-IC PPA?

Even though system partitioning for 3D-ICs is not a novel idea~\cite{yan_how_2006}, a fully native 3D flow, one that would enable a system designer to work directly in 3D, does not exist yet. In the recent years, several pseudo-3D flows emerged from academic research as an extension to the traditional P\&R environments, leveraging existing commercial tools to pile up layers of gates.
Among the most recent, we can cite (1) \texttt{TP-GNN}~\cite{lu_tp-gnn_2020}, (2) \texttt{Pin-3D}~\cite{pentapati_pin-3d_2020}, and (3) \texttt{Snap-3D}~\cite{vanna-iampikul_snap-3d_2021}.\\
(1) \texttt{TP-GNN} leverages routing information to deduce the weights of the corresponding hypergraph.
The hypergraph is then translated into a graph using the k-clique model, despite the loss of partitioning quality highlighted by~\cite{IhlerEdmund;WagnerDorothea;Wagner1993}.
Their Graph Neural Network (GNN) aggregates the graph features to keep together cells that are close to each other and highly interconnected.
The final graph partitioning step uses a k-means clustering with only two centroids to generate two tiers on which to spread the cells, without taking into account the inter-tier connectivity as an objective.\\
(2) \texttt{Pin-3D} feeds upon the partitioning from either \texttt{Shrunk-2D}~\cite{Panth2017} or \texttt{Compact-2D}~\cite{Ku2020}, but replaces their legalization steps with its own 3D-aware pin placement manipulation.
The partitioning is done first, on a regular 2D placement of the circuit, using the Fiduccia-Mattheyses algorithm~\cite{Fiduccia1982} to operate a min-cut on bins of cells. This is a scheme where the floorplan is split into packs of cells according to a grid, then each bin is split individually. It comes down to a gate-level partitioning where we don't allow the partitioner to actually look at the full picture.\\
(3) \texttt{Snap-3D} works by halving the cells height and the floorplan area, then placing the cells using rows and a commercial 2D placer. Then all the cells are restored to their original size and to avoid overlapping, each other row is moved to another tier, creating a 3D arrangement.
The same partitioning scheme as \texttt{Pin-3D} is used.

Recent developments have also been made toward native 3D placement tools, such as \texttt{ART-3D}~\cite{murali_art-3d_2022} that breaks free from tier partitioning. If those research interests certainly foreshadow general progress in the field, they are walking a different path from the present work.
In particular, \texttt{ART-3D} focuses on reducing the critical path length, rather than the total wire-length, by optimizing each individual wires. 

\texttt{Macro-3D}~\cite{Bamberg2020} deserves a mention as another flow proposition that also targets Face-to-Face (F2F) 3D-ICs. By leveraging 2D information and commercial 2D implementation flows, it specifically targets Memory-on-Logic arrangements. In this case, partitioning is not of the essence, as all the memory macros are placed on the top die and the remaining logic on the bottom. The wonders it plays rather regard the placement of the bottom logic relative to the pins of the top macros.

One of the limitations of the presented methods (\texttt{ART-3D}, \texttt{Pin-3D} and \texttt{TP-GNN}) is that they mainly target monolithic 3D-ICs (M3D), where two Front End of Line (FEOL) are placed very close, one on the top of the other, resulting in intimate and fast gate-to-gate connections~\cite{9492421}. As such they cannot be applied to 3D stacking in F2F arrangement where two pins on bottom and top dies need to span two Back End of Line (BEOL) stacks plus the 3D interface, with a non negligible total RC. For such 3D technologies we need a more careful approach regarding the selection of 2D wires that should become 3D, i.e. 3D partitioning. 
\texttt{Snap-3D} and \texttt{Cascade2D} can both target F2F ICs, but their limitation lies in the partitioning itself, as we will discuss in Section~\ref{sec:problem-statement}.

\section{Problem Statement and Contributions}\label{sec:problem-statement}

The existing 3D design flows are using a placement-driven partitioning~\cite{Panth2015} that mainly works in two steps: (1) form regular squares ``bins" of cells and (2) bi-partition each bin using a modified Fiduccia--Mattheyses algorithm.
Such a processing comes down to a gate-level partitioning, each individual gate being open to be moved to another layer, with the objective of limiting the amount of 3D connections in each bin only.
This local minimization of 3D connections does not necessarily translate into a global minimal amount of those connections. If this is fine for M3D, we need to be more careful about those resources with F2F. The technology gap is narrowing thanks to hybrid F2F bonding, enabling a 3D pitch of less than 1µm and allowing very dense die-to-die interconnect at zero area overhead, but a gap there still is.

The proposed solutions are stiff in their own way; they treat all circuits the same way without any tailoring to the specificity of different existing architectures.
There is no leeway built into the flow to let a designer, or the tool itself automatically, use a different partitioning scheme that might adapt better to one design or another.
The focus of those flows is set on post-routing optimizations, rather than on the partitioning itself, which is commendable to an extent.
Our argument is that having a flexible partitioning decision, rather than a ``one size fits all'' with a set in stone method, would only improve those 3D-ICs.

\begin{figure}
\centering
\includegraphics[width=\linewidth]{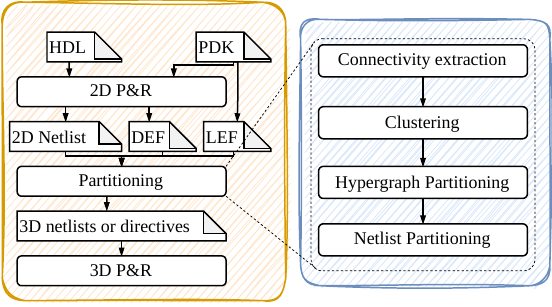}
\caption{Pseudo-3D flow (left) and our automated partitioning (right).}
\label{fig:toolchain}
\end{figure}

To fulfill that objective, we developed an automated partitioning flow that can accommodate for different clustering and partitioning methods, as to allow the user to test which one is the most appropriate for the design considered.
It works in four steps illustrated on the right of Figure~\ref{fig:toolchain}: (1) The connectivity is extracted from the post-placement geometry stripped of its buffer tree and with its wire-length approximated. (2) The resulting hypergraph is clustered to preserve features such as local connectivity. (3) The third step operates a balanced bi-partitioning on the hypergraph using known min-cut algorithms. (4) The partitioning decision associates each gate with a target die and generates netlists or directives that can be fed to the subsequent pseudo-3D P\&R.\\
In the end, we rely on the initial 2D placement to indicate our flow if functional blocks are highly interconnected and close to each other. If they check both boxes, the initial clustering and partitioning will most probably keep them together on the same die to cut fewer nets and not cut short nets.

Beyond having an automated and flexible partitioning decision, the question of which version of the design to partition is often overlooked.
The flows presented in Section~\ref{sec:intro} partition the circuit as-is, considering all wires equal.
Not all 2D wires should be turned into 3D wires, though. Local connections should be preserved to ensure that we do not worsen their performance, in particular in a F2F 3D stacking arrangement where 3D wires need to traverse two full BEOL.
Therefore a preliminary clustering of the circuit can help steering the partitioning into cutting appropriate wires that would actually benefit from a 3D route.

Furthermore, existing flows assume that the single method implemented is universally superior, whatever the design fed to it. Our argument is that a flow with interchangeable blocks would take better benefits of 3D. Allowing the designer to actually choose a different clustering or partitioning method is our way to accommodate for design features.\\
~\\
We bring the following contributions in this paper:
\vspace*{-1em}
\begin{itemize}
\setlength\itemsep{-0.5em}
    \item A flexible and automated partitioning flow taking a 2D circuit and producing 3D netlists and directives compatible with pseudo-3D P\&R flows (Die-by-Die or concurrent 3D placement).
    \item A comparison of 3D PPA gains against the original 2D for several designs.
    \item The analysis of the impact of a preliminary clustering on the partitioning decision quality and 3D WL savings.
    \item A progressive-wire-length clustering method
\end{itemize}

\section{Clustering Methods Considered}\label{sec:clustering-methods}
We will present hereunder three different methods whose impact on the partitioning for 3D decision will then be evaluated.
Hierarchical Geometric (akin to a divisive hierarchical method~\cite{Rokach2005}) and K-means~\cite{Lloyd1982} are known geometrical clustering methods.
Progressive-wire-length, similar in some way to an agglomerative method~\cite{Rokach2005}, is to the best of our knowledge a novel method that has not been used before to study 3D-ICs partitions. It systematically absorbs short wires to hide them inside clusters. 
All the proposed methods are based on geometric information, such as the WL of the nets.
The capacitance of a wire being proportional to its length, focusing on reducing the total system WL is akin to reducing the total power consumption.

\subsection{Hierarchical Geometric (HG)}
\begin{figure}
\centering
\includegraphics[width=.8\linewidth]{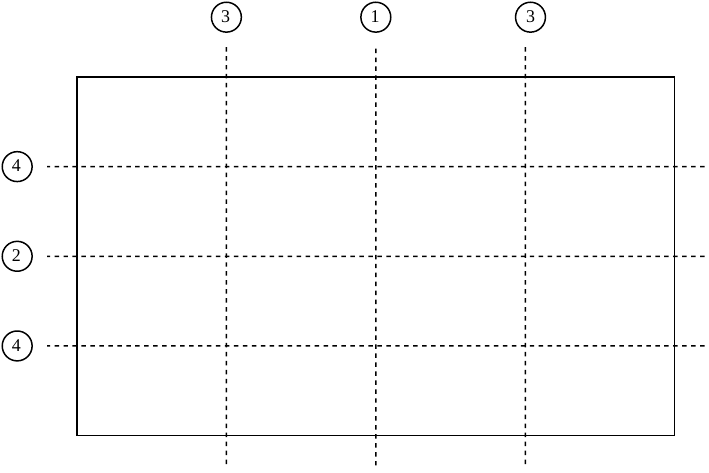}
\caption{Hierarchical geometric method with successive vertical and horizontal slices to create the clusters. Instances overlapping a border are fixed in the cluster in which they have their origin.}
\label{fig:hg}
\end{figure}

We can create clusters in a hierarchical manner by geometrically splitting the design vertically and horizontally successively until the target amount of clusters is reached (illustrated on Figure~\ref{fig:hg}).
Using this method allows to compare different clustering grains as the different levels all share common borders.
The amount of inter-cluster wires (as opposed to \textit{intra}-cluster wires that are entirely contained inside a single cluster) will be growing with the number of clusters.
Finding an optimum  or a trade-off will not depend on finding a magic number, but rather compromising on the average length of inter-cluster WL (which we want high) against the amount of inter-cluster wires (which we want low), as to expose longer nets.

\subsection{K-means (K-m)} 
\begin{figure}
\centering
\includegraphics[width=.8\linewidth]{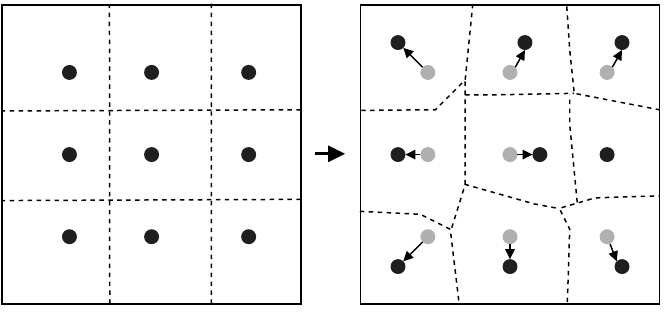}
\caption{K-means method where seeds (dots) are first spread regularly through the floorplan, dashed lines representing the cluster borders gathering around them. Seed positions are then updated to follow the center of gravity of their cluster.}
\label{fig:kmeans-geom}
\end{figure}

K-Means is a method that creates clusters by aggregating elements around the closest seed, with a distance metric based on Manhattan's length in our implementation, which will build a Voronoi diagram of the design using an iterative refinement (illustrated on Figure~\ref{fig:kmeans-geom}).

In our case, the implementation begins by placing $k$ seeds $s_1^{(1)},...,s_k^{(1)} \in \mathbb{R}^2$ regularly spread in the design, around which $k$ clusters $C$ will be grown.
They will act as centers of gravity before starting the first iteration.
We then update the seeds positions by computing the center of mass of each cluster by averaging the position $c_p$ of all the cells it contains:
$$s_i^{(t+1)} = \frac{1}{\left|C_i^{(t)}\right|} \cdot \sum_{c_p \in C_i^{(t)}} c_p$$
We end the iteration by checking if the newly computed seeds $s_i^{(t+1)}$ are close enough to the old $s_i^{(t)}$ ones and if not, start a new iteration with the new seeds until actually reach convergence.
Iteration $t$ starts with the assignation of standard cells to their cluster.
For each standard cell $c$, we find the closest seed $s$ and gather around it:
$$C_i^{(t)} = \left\{c : ||c_p - s_i^{(t)}|| \leq ||c_p - s_j^{(t)}|| \forall j, j \in [1,k]\setminus \{i\}\right\}$$
where $c_p \in \mathbb{R}^2$ is the center of each standard cell.

Ideally, we would let the algorithm run until their is no difference any more for each cluster, but the extreme convergence point might never be reached, and even in the case it could be reached, the computation time would drastically increase compared to a more conservative approach.
We thus arbitrarily chose the convergence criteria as at least 95\% of the seed updating being at most 1 average-gate-width (agw), which is the average width of the all the standard cells used in the circuit:
$$P_{95}\left(||s_i^{(t+1)} - s_i^{(t)}||\right) < 1 \cdot agw$$
This method has the advantage to be able to set an arbitrary amount of clusters, allowing for a fine refinement of the clustering grain.

\subsection{Progressive-wire-length (P-WL)} 
\begin{figure}
\centering
\includegraphics[width=\linewidth]{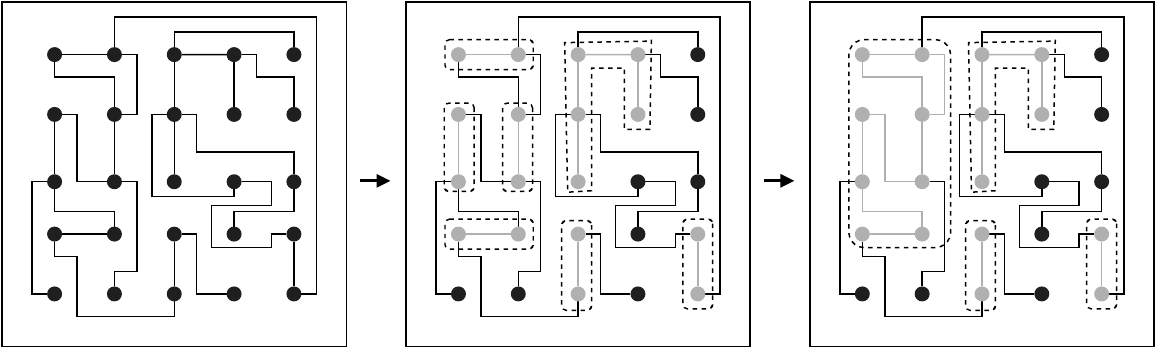}
\caption{Progressive-wire-length method. The first panel is a mock circuit with instances (dots) interconnected. In the second panel, the instances connected by the shortest wire-length were merged inside a cluster and grayed out. The third panel is one step further, showing how clusters themselves are merged when they are connected by short enough wires.}
\label{fig:p-wl}
\end{figure}

The purpose of this method is to absorb all the shortest net into clusters, up to a prefixed length threshold.
Initially, each instance in placed in its own cluster of one element.
Then each connected net is analyzed and if its length is shorter than a threshold, all the clusters it interconnects are merged into a single cluster.
Progressively, all short nets will be absorbed into clusters, effectively hiding them from the partitioning algorithm (illustrated on Figure~\ref{fig:p-wl}).
Let a set of standard cells $c$, a net $n_{ij}$ interconnecting the cells $c_i \in c$ and $c_j \in c$ and the WL $w_{ij}$ of net $n_{ij}$.
If our WL threshold is $t$, a cluster $C_k$ would be defined as
$$C_k = \left\{c_i : \exists n_{ij} \wedge w_{ij} < t \wedge c_j \in C_k, \forall c_i \in c\right\}$$

The downsides are twofold: (i) we might be absorbing longer nets as well into the clusters, preventing them from optimization through 3D routing and (ii) this method actually grows a few very large clusters that may impede a balanced bi-partitioning.
However, in the former case, it is a trade-off between losing performance on short nets being elongated \textit{versus} gains by shortening longer wires.
In the later case, having very large clusters simply means that the threshold length to absorb wires has a design-based hard-limit that will limit the adequacy of the clustering method to some designs.

\textit{In fine}, the progressive-WL method preserve all the shortest connections up to a chosen point.
This ensures that all those short nets cannot be cut by the partitioning, preventing them the need to be routed through the full BEOL to span both dies if indeed cut.
Local connections should stay local as much as possible and not made 3D. This clustering method aims specifically at that objective.
The arbitrary length threshold is set such that the clustering merger does not reach a point where a balanced bi-partitioning is not achievable anymore.

\section{Experimental Setup \& Physical Implementation}\label{sec:setup}
We used two synthetic designs to highlight the impact of the clustering method on the partitioning quality. They both are an assembly of 16 Low-Density Parity-Check (LDPC) cores either daisy chained (LDPC-4x4 Serial, or L4S) or all interconnected to each other in a one-to-all fashion (LDPC-4x4 Full, or L4F).
One LDPC core has a 1024-bit input bus and a 1024-bit output bus. In L4S, core $n$ connects all of its output to the inputs of core $n+1$. In L4F, the output bus of core $n$ is split into 15 and each part is connected to one of the fifteen other cores.
They take place at both ends of the interconnect spectrum: L4S should represent circuits with a high amount of local connectivity and fewer global nets, whereas L4F, due to its fully interconnected nature, should embody a denser intermediate and global interconnects. In a 2D implementation, L4F would not be deemed an appropriate option as the design utilization (DU) ends up being critically low to achieve a low amount of Design Rules Violations (DRVs). By studying a classical approach (L4S) and a highly interconnected one (L4F), we want to show how 3D can enable system designer to consider new ways to interconnect blocks in their architectures.
Specific use-cases, such as MemPool-3D~\cite{cavalcante_mempool-3d_2022}, have been shown to be limited by 2D routing congestion and then improved by a 3D implementation.

The 2D synthesis and P\&R has been accomplished using Cadence Genus and Innovus 21.15 tools with and advanced CMOS predictive PDK from imec, but note that our partitioning only requires instance sizes from the PDK, everything else is handled from the placement geometry. The methodology aims at a minimal die area, or a maximum DU, to minimize the total design wire-length. A maximum DU is one that provides an acceptable amount of DRVs, set to fewer than 1,000 in these experiments, in order for a semi-automated DRV fixing to remain reasonable. We ended up with 97 DRVs for L4S and 531 for L4F. The target clock period has been fixed so that after post-routing optimization, the timing analysis still exhibits some Worst Negative Slack (WNS). However, the timing constraints have been fixed so that the WNS does not exceed 10\% of the target clock period, which is set to 0.9ns in this experiment. The effective operating frequency has been reported as the difference between the target period and the WNS after the post-routing optimization stage. We have used a Power Delivery Network (PDN) with properties (Vdd and Vss lines pitch) that have been already validated in terms of IR-drop.

For 3D system integration we have been using the Concurrent 3D flow from Cadence, part of their Integrity 3D-IC solution, introduced only recently. The idea behind the flow is very similar to the one shown in \texttt{Cascade2D}~\cite{Chang2016}. The main advantages of the flow compared to a legacy Die-by-Die flow are: a) 3D dies are now in the same design data base; b) netlist partitioning and 3D timing constraints are automatically derived from the 2D input netlist and 2D timing constraints; c) standard placement, clock tree synthesis and routing can be used as such, together with their optimization strategies; d) 3D structures placement is fully automated an co-optimized between the two dies; e) once the system is routed, each die data is automatically generated for use with any legacy 3D sign-off tools (e.g. power, thermal, and EM simulation). 
This new flow considerably simplifies 3D systems implementation compared to a legacy Die-by-Die flow, for which we had to provide not only separate netlists for each die, but also timing constraints on a per die basis (input and output delay for 3D nets). 
Partitioning directives from the proposed flow have been directly integrated in the flow. 

As stated earlier in this paper, we assumed a Face-to-Face Wafer-to-Wafer bonding, with a 3D pitch of 0.56µm. This is quite aggressive from 3D technology perspective, but necessary due to 3D net count involved in the design and overall die area. In the following runs, we have not implemented the 3D PDN, the 3D interface is thus composed of signal nets only. A significant amount of the remaining 3D structures could be used for power routing, as each die does implement a 2D PDN (not routed, though).

For both geometrical clustering methods, a grain of 1000 clusters has been selected, as it displays a profitable ratio of intra-cluster wires \textit{versus} inter-cluster wire-length~\cite{Delhaye2019}. As for the progressive-wire-length method, the target maximum WL to be absorbed inside clusters has been set so that the resulting Min-cut partitioning could still be balanced, which came to be 100 times the average-gate-width of the design. Above that threshold, a single cluster is larger than half the area, making it impossible to obtain a balanced bi-partition area-wise.

For the partitioning, the hyperedge weight were set to minimize the amount 3D wires.
We used \texttt{hMetis}~\cite{Karypis1999} version \texttt{2.0-pre1} with an unbalance target of at worst 49/51.
Note that before processing the 2D circuit, they have been stripped from their buffer tree as to remove as much as possible 2D-specific optimizations that would be removed anyway during the 3D P\&R. Keeping those buffers would only have mislead the partitioning decision by shortening buffered, and thus critical, nets and might have led to imbalanced partitions with an uneven buffers distribution across the layers.


\section{Results}

The results will be approached in two steps: the post-partitioning results that give us first insights, then the post-implementation PPA analysis.

\subsection{Post-partitioning analysis}

\begin{figure}[]
\subfloat[L4S\label{fig:netcut-L4S}]{%
    \includegraphics[width=.5\linewidth]{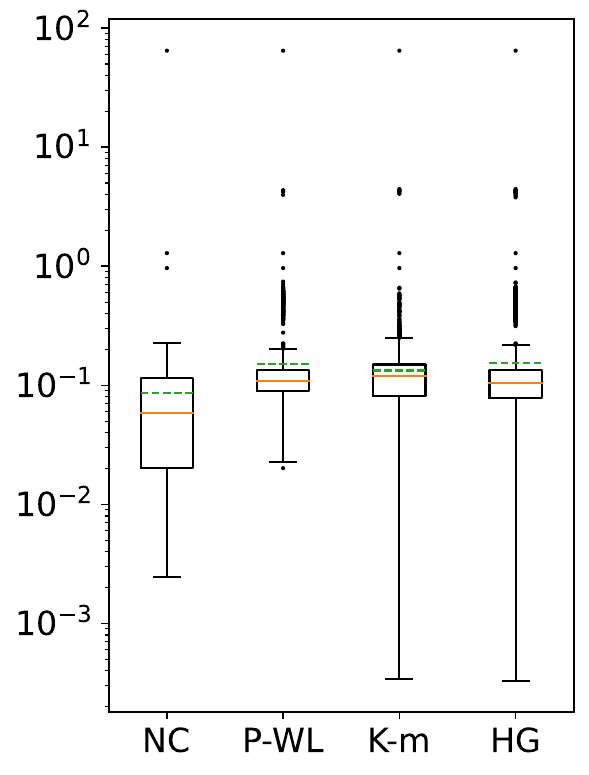}
}
\subfloat[L4F\label{fig:netcut-L4F}]{%
  \includegraphics[width=.5\linewidth]{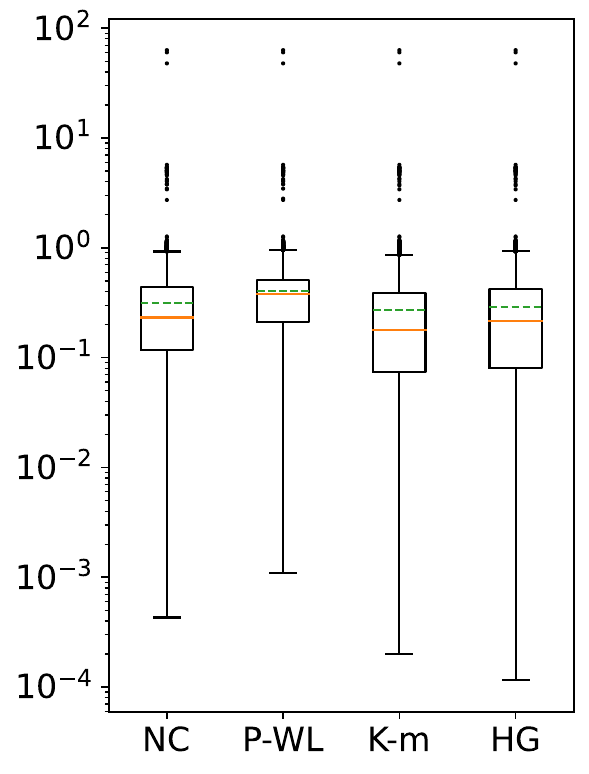}
}
\caption{Distribution of length of cut wires. All WL are normalized on their respective design half-perimeter length. The plain line is the median and the dashed line the average.}
\label{fig:netcut}
\end{figure}

\begin{table}[]
\renewcommand{\arraystretch}{1.2}
\centering
\caption{Post-partitioning cut results. The less nets cut, the better. The higher total WL cut (expressed as a percentage of the total system WL), the better.}
\label{tab:part-results}
\footnotesize
\begin{tabular*}{.85\linewidth}{lcccc}
\toprule
& \multicolumn{2}{c}{L4S} & \multicolumn{2}{c}{L4F} \\
 \cmidrule(lr){2-3} \cmidrule(lr){4-5}
& Nets cut & Total WL cut & Nets cut & Total WL cut \\ 
\cmidrule(lr){2-2} \cmidrule(lr){3-3} \cmidrule(lr){4-4} \cmidrule(lr){5-5}
NC            & \textbf{4,107}  & 3.5\%          & \textbf{14548} & 31.5\%          \\
P-WL          & 4,377           & 6.5\%          & 14680          & \textbf{40.9\%} \\
HG            & 5,575           & \textbf{8.5\%} & 16465          & 32.8\%          \\
K-m           & 5,316           & 6.9\%          & 16388          & 30.4\%          \\
\bottomrule
\end{tabular*}
\end{table}

Looking closer into the resulting partitioning prior to 3D placement, we can analyze the distribution of the WL of the wires cut, as shown on Figure~\ref{fig:netcut}. The aim being to reduce the total system wire-length, a good partitioning will cut longer wires.
Focusing on L4S first (Figure~\ref{fig:netcut-L4S}), not applying any clustering prior to partitioning results in nets significantly shorter than the other three methods. Table~\ref{tab:part-results} highlights the same observation by showing a total wire-length cut only half of the clustering methods, for a similar amount of nets cut.
Applying the different clustering methods seems to yield similar partitions, save for P-WL which cannot cut as shorts wires by design.
On the L4F side (Figure~\ref{fig:netcut-L4F}), applying no clustering (NC) seems to play similarly to the two geometric clustering methods (K-m and HG), albeit cutting less wires (see Table~\ref{tab:part-results}). Yet, P-WL takes a comfortable lead in total WL cut whilst cutting less wires than the other clustering methods.

Furthermore, Table~\ref{tab:part-results} shows us that if we want to minimize the 3D nets count, not applying any clustering is always better. This rightly makes sense as we apply a Min-cut on the whole hypergraph without hindrance. However, by allowing some margin on the cut-size by using a preliminary clustering, we can cut a higher total WL and open the partition to more WL gains.

These first results indicate several insights: (1) There is not a single clustering method that systematically outclasses its competitors, P-WL is a strong contender but HG manages to cut a higher total WL cut for L4S; (2) Not applying a preliminary clustering (NC) allows for fewer nets cut (if only by a small margin) but at the cost of cutting shorter wires, thus encouraging  the use of clustering to guide partitioning.\\

\subsection{Post-implementation analysis}

\begin{table*}[]
\renewcommand{\arraystretch}{1.2}
\centering
\caption{3D physical implementation reports. Percentages give the relative difference with 2D. Power has been reported using statistical toggle rate of the input. P-WL\textsuperscript{\textdagger} is a 3D implementation with the same (limited) DU as 2D. All other implementations are so that the DU is the best possible within the other constraints.}
\label{tab:3D-results}
\footnotesize
\begin{tabular*}{\linewidth}{clccccccccc} \toprule
              & & DRV & Eff. Freq. [GHz] & Tot. Pow. [mW] & Cells & Cells Area [µm²] & Footprint [mm²] & DU & Tot. WL [m] & 3D nets\\ 
\cmidrule(lr){3-3} \cmidrule(lr){4-4} \cmidrule(lr){5-5} \cmidrule(lr){6-6} \cmidrule(lr){7-7} \cmidrule(lr){8-8} \cmidrule(lr){9-9} \cmidrule(lr){10-10} \cmidrule(lr){11-11}

\parbox[t]{2mm}{\multirow{4}{*}{\rotatebox[origin=c]{90}{L4S}}} 
& NC      & 76  & 1.11 (+0.1\%)           & 285 (-4.4\%)          & \textbf{523k (-0.6\%)} & \textbf{16,958 (+1.4\%)} & 11,236 (-48\%) & 74.4\%          & 2.9 (-2.2\%)          & \textbf{4.1k} \\
& P-WL    & 224 & \textbf{1.12 (+1.2\%) } & \textbf{279 (-6.4\%)} & 534k (-0.1\%)          & 16,969 (+1.5\%)          & 11,236 (-48\%) & 75.5\%          & \textbf{2.8 (-4.5\%)} & 4.3k          \\
& HG      & 527 & 1.09 (-1.7\%)           & 281 (-5.7\%)          & 534k (+1.5\%)          & 17,104 (+2.3\%)          & 11,236 (-48\%) & \textbf{76.1\%} & 2.9 (-2.5\%)          & 5.6k          \\
& K-m     & 788 & 1.09 (-2.1\%)           & 287 (-3.7\%)          & 528k (+0.4\%)          & 17,040 (+1.9\%)          & 11,236 (-48\%) & 75.8\%          & 3.0 (+0.5\%)          & 5.3k          \\

\midrule

\parbox[t]{2mm}{\multirow{4}{*}{\rotatebox[origin=c]{90}{L4F}}} 
& NC      & 345 & 1.06 (+8.5\%)         & \textbf{303 (-22\%)} & 578k (+4.0\%)          & \textbf{17,072 (-0.3\%)} & 11,289 (-79\%) & 75.6\%          & 3.7 (-28\%)          & 17.9k          \\
& P-WL    & 739 & \textbf{1.12 (+15\%)} & 307 (-21\%)          & \textbf{563k (+1.3\%)} & 17,768 (+3.3\%)          & 11,289 (-79\%) & 78.3\%          & \textbf{3.6 (-31\%)} & \textbf{15.9k} \\
& P-WL\textsuperscript{\textdagger}  & 314 & 1.10 (+13\%) & 313 (-19\%) & 572k (+2.9\%) & 17,748 (+3.6\%) & 15,438 (-49\%) & 57.5\% & 4.3 (-18\%) & 15.9k \\
& HG      & 785 & 1.06 (+8.5\%)         & 317 (-18\%)          & 594k (+6.9\%)          & 17,910 (+4.6\%)          & 11,289 (-79\%) & 79.3\%          & 3.9 (-25\%)          & 19.6k          \\
& K-m     & 895 & 1.09 (+11\%)          & 310 (-20\%)          & 586k (+5.4\%)          & 18,019 (+5.2\%)          & 11,289 (-79\%) & \textbf{79.8\%} & 3.9 (-25\%)          & 19.8k          \\

\bottomrule
\end{tabular*}
\end{table*}

\begin{table}[]
\renewcommand{\arraystretch}{1.1}
\centering
\caption{Qualitative comparison of the applied clustering methods.}
\label{tab:clustering-methods}
\footnotesize
\begin{tabular*}{\linewidth}{lp{.39\linewidth}p{.39\linewidth}}\toprule
& \multicolumn{1}{c}{L4S} & \multicolumn{1}{c}{L4F} \\
\cmidrule(lr){2-2} \cmidrule(lr){3-3}
\multirow{3}{*}{P-WL} & \tabitem Better performance & \tabitem Better performance \\
                      & \tabitem Better power       & \tabitem Fewest 3D nets \\
                      & \tabitem Better total WL    & \tabitem Lowest total WL \\
\midrule
\multirow{2}{*}{HG}   & \tabitem Slightly worse performance and power & \tabitem Same performance as NC\\
                      & \tabitem Slightly higher DU   & \tabitem Worse power, total WL and 3D nets\\
                      & \tabitem Worst 3D nets        & \\
\midrule
\multirow{2}{*}{K-m}  & \tabitem Worst power      & \tabitem Better performance than NC\\
                      & \tabitem Worst total WL   & \tabitem Worse power, total WL and 3D nets\\
\bottomrule
\end{tabular*}
\end{table}

In Table~\ref{tab:3D-results} we have post-3D placement metrics,  which allows us to bring a grain of salt to the first part of the analysis.
First and foremost, applying a preliminary clustering before partitioning the system is beneficial for almost all metrics.

If we focus on L4S first, what should jump at the reader is that 3D offers very limited gain, when it does not even degrade the PPA. As we presented the design in Section~\ref{sec:setup}, all its functional blocks are connected in a serial fashion; each only sees its one or two neighbors. As such, splitting the design in two parts and stacking them does not much influence the congestion. There are actually very few long nets that require shortening.
Nonetheless, the 3D implementation allows for half the footprint and somewhat reduces the total power dissipation.

We can still spot differences in the preliminary clustering, such as the way that NC does create the least amount of 3D wires that account for 5.8\% of the total 3D structures for the 3D pitch and footprint considered.
However, by using 0.3\% more 3D structures after applying a P-WL clustering, we can gain on total system WL, reaching 4.5\% savings.
The design utilization stays in the same range as the 2D implementation that was at 77.4\%.\\

L4F is a interesting case by itself due to its highly dense interconnect.
Its 2D DU can only climb to 57\% due to the congestion in the BEOL if we want to keep the DRV below 1,000 as stated in Section~\ref{sec:setup}. This is a typical case of a design that would be overlooked if the implementation target was a 2D circuit.
Going 3D relaxes the congestion greatly and allows for a denser cell placement that results in a 3D DU between 75\% and 80\%, figures in the same ballpark as its L4S cousin.

\begin{figure}[]
\subfloat[L4S NC\label{fig:L4S-01-3D}]{%
    \includegraphics[width=.22\linewidth]{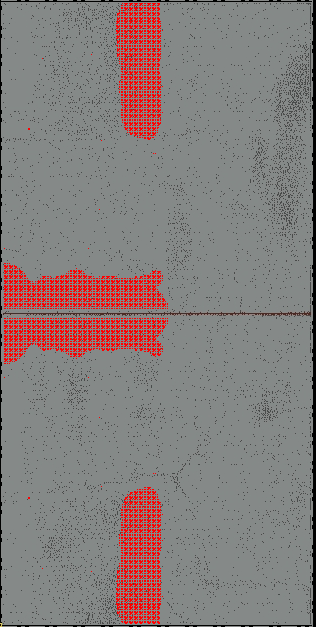}
}
\subfloat[L4S P-WL\label{fig:L4S-04-3D}]{%
  \includegraphics[width=.22\linewidth]{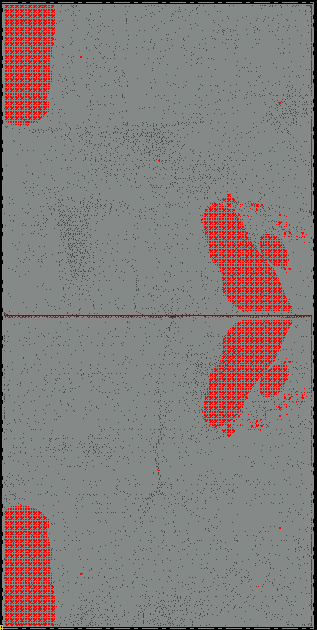}
}
\subfloat[L4S HG\label{fig:L4S-02-3D}]{%
  \includegraphics[width=.22\linewidth]{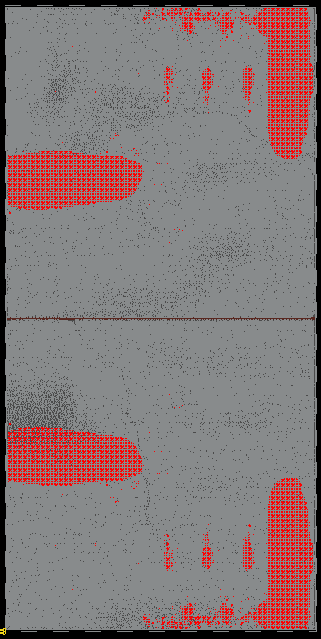}
}
\subfloat[L4S K-m\label{fig:L4S-03-3D}]{%
  \includegraphics[width=.22\linewidth]{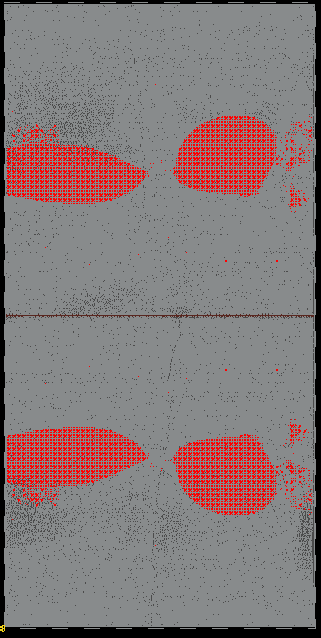}
}\\
\subfloat[L4F NC\label{fig:L4F-01-3D}]{%
    \includegraphics[width=.22\linewidth]{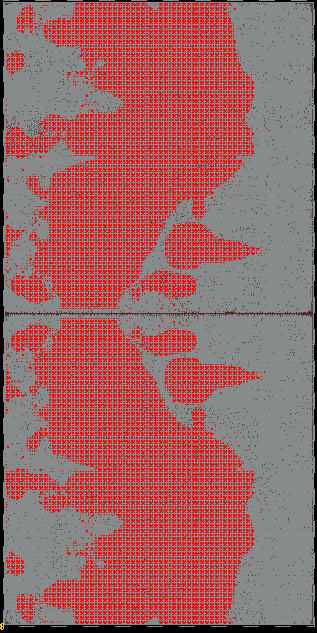}
}
\subfloat[L4F P-WL\label{fig:L4F-04-3D}]{%
  \includegraphics[width=.22\linewidth]{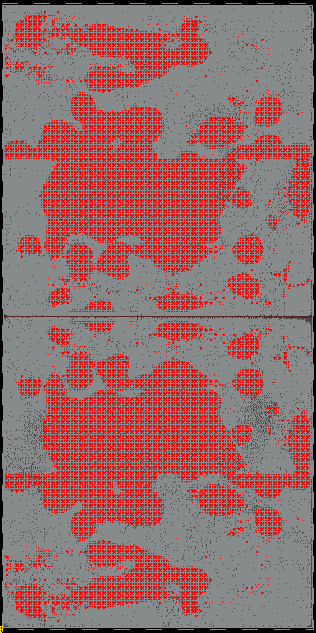}
}
\subfloat[L4F HG\label{fig:L4F-02-3D}]{%
  \includegraphics[width=.22\linewidth]{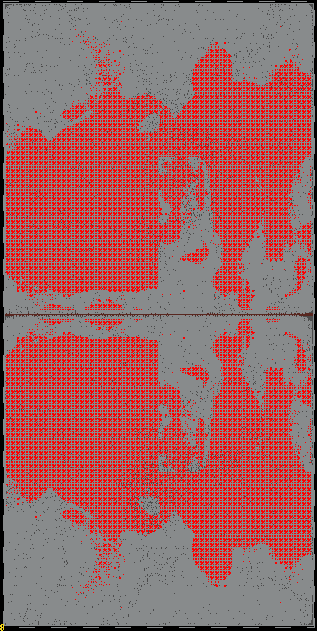}
}
\subfloat[L4F K-m\label{fig:L4F-03-3D}]{%
  \includegraphics[width=.22\linewidth]{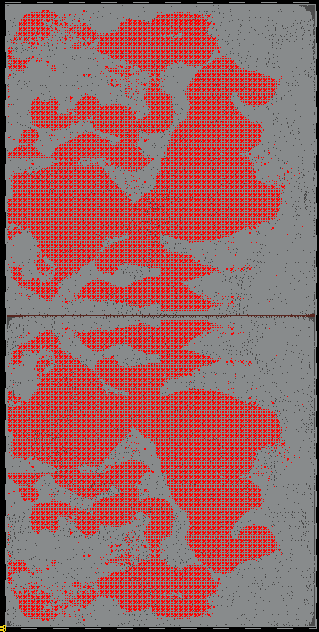}
}
\caption{Physical implementation of all designs, with top and bottom 3D structure allocation highlighted (one die is flipped with respect to the other).}
\label{fig:3D}
\end{figure}

Across the board, we get significant gains in power, performance and footprint area.
Looking at the 3D nets, we can spot a slight discrepancy with the post-partitioning results, all methods actually cutting more wires than anticipated. This is especially noticeable for NC that now populates 25\% of the 3D structures, compared to 22\% for P-WL. This might be explained by extra buffers insertion and 3D connections in NC to comply with timing constraints.
Figure~\ref{fig:3D} illustrates the stark difference in 3D structures utilization between L4S (top row) and L4F (bottom row).
Even if L4F uses up to 28\% of its budget with K-m and HG, there are still enough pins available for PDN routing.
Regarding power consumption, HG performs noticeably worse than its contenders. NC keeps a slight edge over P-WL, when the situation was reversed for L4S.
At the end of the day, P-WL is still the most performant by a comfortable relative margin.
Table~\ref{tab:clustering-methods} summarizes the qualitative comparison of the clustering methods applied prior to partitioning both designs.

Instead of pushing the 3D implementation to its limits with a maximum DU, we could have limited it to a more traditional iso-DU implementation by halving the 2D footprint.
Our aim in this experiment is to see how much a highly congested interconnect such as L4F can benefit from a 3D transition, but let's take the other point of view for a moment.
The P-WL implementation of L4F has been implemented with a halved footprint in the same Table~\ref{tab:3D-results}. The performance are slightly worse, the power consumption slightly higher, and the total wire-length significantly increased due to the spread of the cells to limit the congestion.
Even by restricting the 3D possibilities, it still outperforms the 2D and makes a viable implementation.
The worst part would be the higher footprint compared to an unrestricted 3D, that would incur increased manufacturing costs.

In the end, depending on the design, our partitioning flow highlights that selecting an appropriate clustering method can yield a difference of up to 31\% in total WL, 22\% in total power and 15\% in effective frequency.
Furthermore, we see that a local interconnect-dominated design such as L4S only marginally benefices from a 3D arrangement with total power gains not exceeding 6.4\% and effective frequency not climbing more than 1.2\% (when it does not decrease). However a global interconnect-dominated L4F wins the day with a design utilization that wastes way less silicon compared to 2D, and benefits from significant WL, frequency, power and area improvements.
This should encourage system designers to turn their attention to designs with highly interconnected blocks, knowing that a 3D implementation would make it viable.

\section{Conclusion}\label{sec:ccl}
In this paper, we reviewed the impact of clustering methods on the partitioning decision for 3D-ICs. We showed that depending on the design, different clustering methods have a significant impact on the partitioning decision and the resulting 3D system quality, through the measurement of the total system WL, effective frequency and total power.
As there is no universal winner, a careful designer ought to consider and compare several clusterings before settling on a specific technique. Depending on the PPA objectives and the technology constraints, one might need a different method. An interesting extension would be to explore a correlation between clustering analysis and post-3D implementation PPA to classify which method is effective against different classes of design.
Furthermore, a design with a lower 2D design utilization is a better candidate for 3D WL savings, showing that 3D-ICs are an excellent opportunity to design architectures with a denser interconnect. In future works, it would be interesting to extend the present study to more designs, as this preliminary study focused on proving that clustering methods have different impacts depending on the design.
Further analysis of 3D partitions and nets cut could also reveal new insights regarding the type of net cut, e.g. their importance in the critical path, which might lead to a better definition of what constitutes a `good partition'.

 \bibliographystyle{elsarticle-num} 
 \bibliography{cas-refs}





\end{document}